\documentclass[useAMS,usenatbib]{mn2e}
\usepackage{txfonts}
\usepackage{natbib}
\usepackage[all]{xy}
\usepackage[dvips]{graphicx}
\usepackage{feynmp}

\def\del#1{{}}

\def\david#1{{#1}}

\onecolumn

\newcommand{\ltsima}{$\; \buildrel < \over \sim \;$}
\newcommand{\lsim}{\lower.5ex\hbox{\ltsima}}
\newcommand{\gtsima}{$\; \buildrel > \over \sim \;$}
\newcommand{\gsim}{\lower.5ex\hbox{\gtsima}}
\newcommand{\bra}{\langle}
\newcommand{\ket}{\rangle}

\newcommand{\dd}{\mathrm{d}}

\newcommand{\ci}{\mathrm{i}}
\newcommand{\vecx}{\bmath{x}}
\newcommand{\geo}{\vecx(\btheta,\chip)}
\newcommand{\vecl}{\bmath{\ell}}
\newcommand{\veclp}{\bmath{\ell}^\prime}
\newcommand{\trace}{\mathrm{tr}}
\newcommand{\dlp}{(\ell-\lprime)}

\newcommand{\chip}{{\chi^\prime}}
\newcommand{\chipp}{{\chi^{\prime\prime}}}
\newcommand{\lprime}{\ell^\prime}
\newcommand{\dirac}{\delta_D}

\title[Born approximation and flexions]
{On the validity of the Born approximation for weak cosmic flexions}
\author[B.M. Sch\"afer, L. Heisenberg, A.F. Kalovidouris and D.J. Bacon]
{Bj{\"o}rn Malte Sch\"afer\thanks{e-mail:spirou@ita.uni-heidelberg.de}$^{1,2}$, Lavinia Heisenberg$^{3,4}$, Angelos F. Kalovidouris$^2$, David J. Bacon$^5$\\
$^1$ Institut d'Astrophysique Spatiale, Universit{\'e} de Paris XI, b{\^a}timent 120-121, Centre universitaire d'Orsay, 91400 Orsay CEDEX, France\\
$^2$ Astronomisches Recheninstitut, Zentrum f{\"u}r Astronomie, Universit{\"a}t Heidelberg, M{\"o}nchhofstra{\ss}e 12, 69120 Heidelberg, Germany\\
$^3$ Institut f{\"u}r theoretische Astrophysik, Zentrum f{\"u}r Astronomie, Universit{\"a}t Heidelberg, Albert-Ueberle-Stra{\ss}e 2, 69120 Heidelberg, Germany\\
$^4$ Universit{\'e} de Gen{\`e}ve, D{\'e}partement de Physique Th{\'e}orique, 24, quai Ernest Ansermet, 1211 Gen{\`e}ve, Switzerland\\
$^5$ Institute of Cosmology and Gravitation, University of Portsmouth, Dennis Sciama Building, Burnaby Road, Portsmouth PO1 3FX, United Kingdom}

\begin{document}
\pagerange{\pageref{firstpage}--\pageref{lastpage}}
\pubyear{2008}
\maketitle
\label{firstpage}
\unitlength = 1mm

\begin{abstract}
Weak lensing calculations are often made under the assumption of the Born approximation, where the ray path is approximated as a straight radial line. In addition, lens-lens couplings where there are several deflections along the light ray are often neglected. We examine the effect of dropping the Born approximation and taking lens-lens couplings into account, for weak lensing effects up to second order (cosmic flexion), by making a perturbative expansion in the light path. We present a diagrammatic representation of the resulting corrections to the lensing effects. The flexion signal, which measures the derivative of the density field, acquires correction terms proportional to the squared gravitational shear; we also find that by dropping the Born approximation, two further degrees of freedom of the lensing distortion can be excited (the twist components), in addition to the four standard flexion components. We derive angular power spectra of the flexion and twist, with and without the Born-approximation and lens-lens couplings and confirm that the Born approximation is an excellent approximation for weak cosmic flexions, except at very small scales. 
\end{abstract}


\section{Introduction} 
Weak gravitational flexions are third order distortions of the images of distant galaxies due to gravitational lensing by the large-scale structure \citep{2002ApJ...564...65G, 2005ApJ...619..741G, 2006MNRAS.365..414B, 2008A&A...485..363S}. Flexions have the potential to probe the profile \citep{2009MNRAS.396.2257L, 2010arXiv1011.3041V, 2011arXiv1101.4407F}, substructure \citep{2007ASPC..379..338R, 2010MNRAS.409..389B} and ellipticity of dark matter haloes \citep{2010arXiv1009.3125E, 2009MNRAS.400.1132H}, improve mass reconstructions \citep{2010arXiv1008.3088E}, or complement cosmic shear studies as they are likely to be less prone to intrinsic alignment effects \citep[for a review of those, see][]{l_review}. They can be measured by a shapelet decomposition \citep{2005IAUS..225...31M, 2005MNRAS.363..197M, 2007MNRAS.380..229M, 2007ApJ...660.1003G} or by the octupole moments \citep{2005NewAR..49...83I, 2006ApJ...645...17I, 2007ApJ...671.1182I, 2007ApJ...660..995O, 2008ApJ...680....1O, 2008MPLA...23.1506U, 2009ApJ...699..143O} of the brightness distribution of the lensed galaxies.

A common approximation in the theory of weak lensing \citep[for reviews, see][]{1992grle.book.....S, 2001PhR...340..291B, 2010arXiv1010.3829B}, in particular in weak cosmic shear, is the Born-approximation in which one integrates the gravitational tidal field (weighted suitably for lensing) along a fiducial straight ray instead of the actual photon geodesic. The Jacobian matrix, which describes the image mapping to first order, is symmetric in this approximation. It becomes asymmetric, however, if corrections from the perturbed geodesic are taken into account, and/or if multiple deflections along the line of sight are considered \citep{1997A&A...322....1B, 1998MNRAS.296..873S}. The influence of perturbed geodesics on the cosmic shear signal was worked out in detail by \citet{2002ApJ...574...19C}, \citet{2006JCAP...03..007S} and \citet{2009arXiv0910.3786K}, who in addition treat the complication that the actual observable in lensing is the reduced shear $g=\gamma/(1-\kappa)$ instead of the gravitational shear $\gamma$. Recently, \citet{PhysRevD.81.083002} pursue a different approach using the Sachs-equation, which describes the infinitesimal changes in the cross section of a light bundle during propagation; they derive corrections to the power spectra due to the perturbed geodesic and identify further relativistic effects in weak lensing image distortions. These studies agree that corrections due to perturbed geodesics are minor, but that vortical patterns in the shear field can be excited, referred to as $B$-modes \citep{1996astro.ph..9149S}. Additionally, rotation of galaxy images can occur \citep[see e.g.][]{2003PhRvD..68h3002H, 2006MNRAS.367.1543P}. The validity of perturbative techniques has been checked against ray-tracing on N-body simulations of the cosmic density field, and found to be well approximated even on nonlinear scales \citep{2000ApJ...530..547J, 2001MNRAS.322..918V, 2008arXiv0809.5035H, 2009A&A...497..335T}. Statistics of Born-approximated shear and flexion fields have been investigated in detail in \citet{2010arXiv1003.5003M} and \citet{2010arXiv1012.3658M}.

In this paper, we aim to generalise these results by examining the effect of dropping the Born approximation, and including multiple deflections, at the order of gravitational flexions, which is significantly more complex. Flexions describe the components of the tensor $\partial_i\partial_j\partial_k\psi$ i.e. the third angular derivatives of the lensing potential $\psi$; the Born approximation forces this tensor to be symmetric, due to the interchangability of any pair of partial derivatives, giving 4 degrees of freedom. The remaining 2 degrees of freedom, coined {\em twist}, can be present as a systematic or can be physically excited by dropping the Born-approximation, which as we will see destroys the symmetry of the tensor, $\partial_i\partial_j\partial_k\psi\neq\partial_j\partial_i\partial_k\psi$. The resulting full set of six flexion-related image distortion modes has been described by \citet{2009MNRAS.396.2167B}. We aim to provide corrections to the spectra of all flexion components due to the perturbed photon geodesic and to quantify the correlation properties of the newly emerging flexion degrees of freedom. While carrying out the computation, we will point out the analogies to gravitational shear. We work in a flat-sky approximation, because gravitational flexions are a small-scale phenomenon.

The paper is organized as follows. After a compilation of the relevant formulae concerning cosmology and structure formation in Sect.~\ref{sect_cosmology}, we introduce gravitational flexions and compute the correction terms due to relaxation of the Born-approximation in Sect.~\ref{sect_lensing}. We also present a diagrammatic representation of the distortion corrections. Angular power spectra of the flexion quantities with their corrections are shown in Sect.~\ref{sect_spectra}, and the main results are summarised in Sect.~\ref{sect_summary}.

\section{cosmology}\label{sect_cosmology}

In this section we introduce the quantities we require in order to calculate cosmological weak lensing statistics. We require a description of both the cosmological background and the matter fluctuations in the Universe, and we need to be able to describe how these fluctuations grow with time.

\subsection{Dark energy cosmologies}
First, we need to describe the background expansion of the Universe. In spatially flat dark energy cosmologies with matter density parameter $\Omega_m$, the Hubble function $H(a)=\dd\ln a/\dd t$ is given by
\begin{equation}
\frac{H^2(a)}{H_0^2} = \frac{\Omega_m}{a^{3}} + \frac{1-\Omega_m}{a^{3(1+w)}},
\end{equation}
with the dark energy equation of state parameter $w$. The value $w\equiv -1$ corresponds to a cosmological constant $\Lambda$. The relation between comoving distance $\chi$ and scale factor $a$ is given by
\begin{equation}
\chi = c\int_a^1\:\frac{\dd a}{a^2 H(a)},
\end{equation}
in units of the Hubble distance $\chi_H=c/H_0$.

The cosmological model used throughout is a spatially flat $\Lambda$CDM cosmology with Gaussian adiabatic initial perturbations in the cold dark matter density field. The specific parameter choices are $\Omega_m = 0.25$, $n_s = 1$, $\sigma_8 = 0.8$, $\Omega_b=0.04$, $w=-1$ and $H_0=100\: h\:\mathrm{km}/\mathrm{s}/\mathrm{Mpc}$, with $h=0.72$. For simplicity, we assume that the lensed galaxies reside at a redshift of $z_s=0.9$, corresponding to the projected median of the redshift distribution of the EUCLID galaxy sample \citep{2008arXiv0802.2522R, 2007MNRAS.381.1018A}.

\subsection{Matter power spectrum}
We also need to describe the density fluctuations in the Universe. The linear cold dark matter (CDM) density power spectrum $P(k)$ describes the fluctuation amplitude of a Gaussian statistically homogeneous and isotropic density field $\delta$, $\bra\delta(\bmath{k})\delta(\bmath{k}^\prime)\ket=(2\pi)^3\dirac(\bmath{k}+\bmath{k}^\prime)P(k)$, and is given by the ansatz
\begin{equation}
P(k)\propto k^{n_s}T^2(k),
\end{equation}
with the transfer function $T(k)$. In low-$\Omega_m$ cosmologies $T(k)$ can be approximated by \citep{1986ApJ...304...15B},
\begin{equation}
T(q) = \frac{\ln(1+2.34q)}{2.34q}\left(1+3.89q+(16.1q)^2+(5.46q)^3+(6.71q)^4\right)^{-\frac{1}{4}},
\label{eqn_cdm_transfer}
\end{equation}
where the wave vector $k=q\Gamma$ is rescaled with the shape parameter $\Gamma$ \citep{1995ApJS..100..281S} describing corrections due to the baryon density $\Omega_b$,
\begin{equation}
\Gamma=\Omega_m h\exp\left(-\Omega_b\left(1+\frac{\sqrt{2h}}{\Omega_m}\right)\right).
\end{equation}
\david{This fit to $T(k)$ is sufficient here for illustrative purposes.} The spectrum $P(k)$ is normalised to the standard deviation of matter fluctuations $\sigma_8$ on a scale $R=8~\mathrm{Mpc}/h$,
\begin{equation}
\sigma^2_R 
= \frac{1}{2\pi^2}\int\dd k\:k^2 P(k) W^2(kR)
= \int\dd\ln k\: \big(\Delta(k) W(kR)\big)^2
\quad\mathrm{with}\quad
\Delta^2(k)\equiv \frac{k^3}{2\pi^2}P(k)
\end{equation}
where $W$ is a Fourier transformed spherical top hat filter function, $W(x)=3j_1(x)/x$, and $j_\ell(x)$ is the spherical Bessel function of the first kind of order $\ell$  \citep{1972hmf..book.....A}.

\subsection{Structure growth}
Finally in this section, we need to describe the growth of structure as time progresses. Linear homogeneous growth of the density field, $\delta(\bmath{x},a)=D_+(a)\delta(\bmath{x},a=1)$, is described by the growth function $D_+(a)$, which is the solution to the growth equation \citep{1997PhRvD..56.4439T, 1998ApJ...508..483W, 2003MNRAS.346..573L},
\begin{equation}
\frac{\dd^2}{\dd a^2}D_+(a) + \frac{1}{a}\left(3+\frac{\dd\ln H}{\dd\ln a}\right)\frac{\dd}{\dd a}D_+(a) = 
\frac{3}{2a^2}\Omega_m(a) D_+(a),
\label{eqn_growth}
\end{equation}
whose solution is $D_+(a)=a$ if $\Omega_m=1$ and remains close to this solution in dark energy cosmologies. Consequently, the spectrum grows according to $P(k,a) = D_+^2(a)P(k)$. For the purpose of this paper we restrict ourselves to this linear growth regime, despite flexion having much higher signal-to-noise on smaller scales. This is due to the computational complexity of the calculations below; however, we will see that the correction features can be easily understood in this regime and quantitatively extrapolated correctly to the nonlinear regime.


\section{gravitational lensing}\label{sect_lensing}

In this section we will show how the lensing quantities up to third order can be calculated, including lens-lens couplings and dropping the Born approximation. We will also introduce a scheme of diagrams that can describe the various lensing correction terms.

\subsection{Weak lensing and the light propagation equations}
\begin{figure}
\begin{center}
\resizebox{0.9\hsize}{!}{\includegraphics{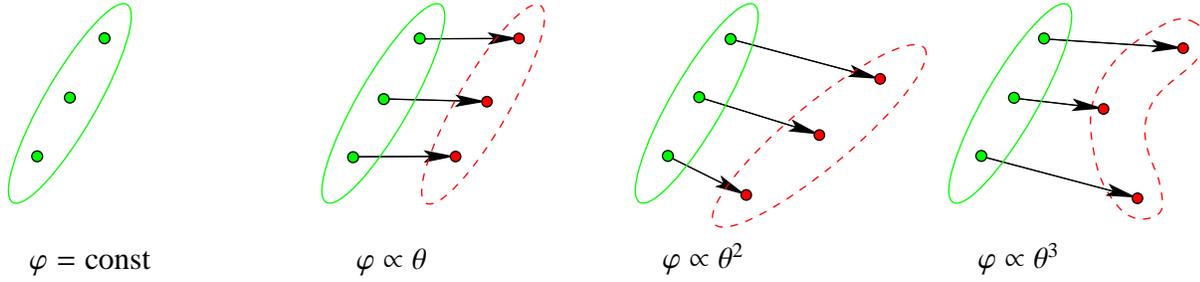}}
\end{center}
\caption{Image distortions induced in gravitational lensing: a constant potential does not affect the image of a galaxy (first panel); a potential varying linearly with position displaces an image (second panel); a potential with second derivatives shears an image (third panel); and non-vanishing third derivatives of the potential bend an image (fourth panel), described by the flexions. Hence, the flexion is related to a quadratic variation of the displacement, or a linear variation of the shear across the galaxy image.}
\label{fig_flexion}
\end{figure}

Weak gravitational lensing occurs when the shapes of distant galaxies are distorted as their light is deflected by intervening gravitational potentials. The light deflection translates the image of a galaxy to a new position on the celestial sphere. If the deflection varies linearly across the galaxy image, it appears sheared. If furthermore the deflection shows an appreciable quadratic variation across the image (i.e. the shear varies linearly with position), one can observe a bending of the image, which is refered to as gravitational flexion. Fig.~\ref{fig_flexion} gives an impression of the image distortions in weak gravitational lensing.

In order to describe these phenomena in detail, we use the common formalism \citep{1991MNRAS.251..600B, 1994CQGra..11.2345S, 1994A&A...287..349S, 1996ASPC...88...57K, 1997ApJ...484..560J, 1998MNRAS.296..873S} for describing light propagation in an inhomogeneous universe: $\vecx(\btheta,\chi)$ is the transverse comoving separation between a light ray and a fiducial ray which intersect at the observer at an angle $\theta$. This separation increases linearly with comoving distance and acquires corrections due to light deflection in gravitational potentials,
\begin{equation}
\vecx(\btheta,\chi) = 
\chi\btheta-\frac{2}{c^2}\int_0^\chi\dd\chip\: (\chi-\chip)\left[\nabla_\perp\Phi(\geo,\chip)-\nabla_\perp\Phi^{(0)}(\chip)\right]
= \chi\btheta-2\int_0^\chi\dd\chip\: (\chi-\chip)\left[\nabla_\perp\varphi(\geo,\chip)-\nabla_\perp\varphi^{(0)}(\chip)\right],
\label{eq:x}
\end{equation}
where we have defined the dimensionless potential $\varphi=\Phi/c^2$ for brevity. Without lensing, a source at comoving distance $\chi$, separated by $\vecx$ from the fiducial ray, will be observed at an angle $\bbeta=\vecx/\chi$. With lensing, this angle becomes
\begin{equation}
\beta_i(\btheta,\chi) = 
\theta_i - 
2\int_0^\chi\dd\chip\:\frac{\chi-\chip}{\chi}\left[\varphi_i(\geo,\chip)-\varphi^{(0)}_i(\chip)\right],
\end{equation}
where the index on the gravitational potential denotes the derivative perpendicular to the line of sight relative to the component $i$ of the physical coordinate. The shear and convergence of galaxy images is related to the Jacobian matrix $\mathcal{A}\equiv\partial\bbeta/\partial\btheta$ of the mapping, which can be obtained by differentiation of the lensed source position $\bbeta$ with respect to the true celestial coordinate $\btheta$ of the source:
\begin{equation}
\mathcal{A}_{ij}(\btheta,\chi) \equiv\frac{\partial\beta_i}{\partial\theta_j} = 
\delta_{ij} - 2\int_0^\chi\dd\chip\:
(\chi-\chip)\frac{\chip}{\chi}\varphi_{ik}(\geo,\chip)\mathcal{A}_{kj}(\btheta,\chip),
\label{eqn_jacobian_los}
\end{equation}
with summation over repeated indices. Note that this will be quite difficult to solve in general, as it is an implicit relation, with $\mathcal{A}$ appearing on both sides. In order to examine flexion-related image shape changes, we require the angular derivative of the Jacobian $\mathcal{A}$ which defines the rank-3 tensor $\mathcal{B}$:
\begin{equation}
\mathcal{B}_{ijk}(\btheta,\chi) \equiv\frac{\partial\mathcal{A}_{ij}}{\partial\theta_k} = 
-2\int_0^\chi\dd\chip\:(\chi-\chip)\frac{\chip^2}{\chi}
\left[
\varphi_{ilm}(\geo,\chip)\mathcal{A}_{lj}(\btheta,\chip)\mathcal{A}_{mk}(\btheta,\chip)+
\frac{1}{\chip}\varphi_{im}(\geo,\chip)\mathcal{B}_{mjk}(\btheta,\chip)
\right].
\label{eqn_djacobian_los}
\end{equation}
It is worth noting that $\mathcal{B}_{ijk}$ is always symmetric under exchange of the last two indices, $\mathcal{B}_{ijk} = \mathcal{B}_{ikj}$, which follows straightforwardly from its definition as the partial derivative of $\mathcal{A}$:
\begin{equation}
\mathcal{B}_{ijk} = \frac{\partial\mathcal{A}_{ij}}{\partial\theta_k} = \frac{\partial^2\beta_i}{\partial\theta_k\partial\theta_j} = \frac{\partial^2\beta_i}{\partial\theta_j\partial\theta_k} = \frac{\partial\mathcal{A}_{ik}}{\partial\theta_j} = \mathcal{B}_{ikj}.
\end{equation}
Due to this symmetry there are in total 6 independent entries in $\mathcal{B}_{ijk}$: $\mathcal{B}_{000}$, $\mathcal{B}_{010}=\mathcal{B}_{001}$, $\mathcal{B}_{011}$, $\mathcal{B}_{100}$, $\mathcal{B}_{101}=\mathcal{B}_{110}$ and $\mathcal{B}_{111}$.

\subsection{Perturbative solution to the optical equations}\label{sect_optical}
The equations for the lensing quantities above can be solved perturbatively: starting with the solutions for the unlensed case (i.e. a straight line, which corresponds to the Born approximation) one recoveres iterative corrections to the geodesic $\vecx(\btheta,\chi)$ and hence for the image position $\bbeta(\btheta,\chi)$, for the Jacobian $\mathcal{A}(\btheta,\chi)$ and finally for the derivative $\mathcal{B}(\btheta,\chi)$. The perturbative expansion is controlled by considering corrections to the geodesic up to $\mathcal{O}(\varphi)$ in the gravitational potential. Consequently, the Jacobian acquires corrections of order $\mathcal{O}(\varphi^2)$ and the derivative is corrected by terms of order $\mathcal{O}(\varphi^3)$.

The photon geodesic is approximated by $\vecx = \vecx^{(0)} + \vecx^{(1)} + \mathcal{O}(\varphi^2)$, with the zeroth order solution $\vecx^{(0)}(\btheta,\chi)$ being a straight line $\vecx^{(0)}(\btheta,\chi) = \chi\btheta$, and a first order correction $\vecx^{(1)}(\btheta,\chi)$ which is linear in the gravitational potential, as seen in equation (\ref{eq:x}):
\begin{equation}
\vecx^{(1)}(\btheta,\chi) = -2\int_0^\chi\dd\chip\:(\chi-\chip)
\left[\nabla_\perp\varphi(\vecx^0(\btheta,\chip),\chip)-\nabla_\perp\varphi^{(0)}(\chip)\right].
\label{eqn_geodesic_first}
\end{equation}
The Jacobian matrix is expanded to second order in the gravitational potential, $\mathcal{A} = \mathcal{A}^{(0)} + \mathcal{A}^{(1)} + \mathcal{A}^{(2)} + \mathcal{O}(\varphi^3)$, and has the unit-matrix as the zeroth-order solution $\mathcal{A}_{ij}^{(0)}=\delta_{ij}$ which corresponds to the unlensed case, i.e. the identity mapping. The first order correction $\mathcal{A}_{ij}^{(1)}$ measures the line of sight integrated tidal fields $\varphi_{ij}$ (the indices on the gravitational potential denote a partial derivative):
\begin{equation}
\mathcal{A}_{ij}^{(1)}(\btheta,\chi) = -2\int_0^\chi\dd\chip\:(\chi-\chip)\frac{\chip}{\chi}
\varphi_{ij}(\vecx^0(\btheta,\chip),\chip),
\label{eqn_jacobian_first}
\end{equation}
and the second order correction $\mathcal{A}_{ij}^{(2)}$ comprises terms caused by lens-lens coupling ($\propto\mathcal{A}_{kj}^{(1)}$) and by the perturbed geodesic ($\propto x^{(1)}_{l}$), following from eqn.~(\ref{eqn_jacobian_los}) and collecting terms quadratic in the gravitational potential, after substitution of eqns.~(\ref{eqn_geodesic_first}) and~(\ref{eqn_jacobian_first}):
\begin{equation}
\mathcal{A}_{ij}^{(2)}(\btheta,\chi) =  
-2\int_0^\chi\dd\chip\:(\chi-\chip)\frac{\chip}{\chi}
\left[
\varphi_{im}(\vecx^0(\btheta,\chip),\chip)\mathcal{A}_{mj}^{(1)}(\btheta,\chip) + 
\varphi_{ijm}(\vecx^0(\btheta,\chip),\chip)x^{(1)}_{m}(\btheta,\chip)
\right],
\label{eqn_jacobian_second}
\end{equation}
where the first-order solutions for the geodesic $\vecx^{(1)}(\btheta,\chip)$ and for the Jacobian matrix $\mathcal{A}^{(1)}_{ij}(\btheta,\chip)$ need to be substituted. The first term is due to lens-lens coupling, as the galaxy image is deformed by the lensing event at $\chip$ and is also deformed by further lensing events in the large scale structure between 0 and $\chip$ via $\mathcal{A}_{mj}^{(1)}(\btheta,\chip)$. The second term is due to the dropped Born-approximation, i.e. the lensing event at $\chip$ picks up the derivatives of the potential at the position $\chip\btheta+\vecx^{(1)}(\btheta,\chip)$ rather than at $\chip\btheta$ because of the perturbed geodesic due to the integrated lensing effect up to the distance $\chip$. These corrections can be interpreted as nonlocal interactions of the light with the gravitational potential and break the symmetry of the Jacobian, because it is no longer simply a projected double derivative of the gravitational potential with interchangeable indices.

Progressing now to the flexion order, the derivative of the Jacobian is expanded to include terms up to third order in $\varphi$, $\mathcal{B} = \mathcal{B}^{(0)} + \mathcal{B}^{(1)} + \mathcal{B}^{(2)} + \mathcal{B}^{(3)} + \mathcal{O}(\varphi^4)$, with the zeroth order solution $\mathcal{B}_{ijk}^{(0)} = 0$, indicating the absence of lensing effects if the light bundle propagates through a constant gravitational potential. The derivative $\mathcal{B}^{(1)}$ measures the third derivative of the gravitational potential,
\begin{equation}
\mathcal{B}_{ijk}^{(1)}(\btheta,\chi) = -2\int_0^\chi\dd\chip\:(\chi-\chip)\frac{\chip^2}{\chi}
\varphi_{ijk}(\vecx^0(\btheta,\chip),\chip).
\label{eqn_flexion_first}
\end{equation}
The higher order corrections are again derived by substituting the perturbative expansion of all relevant terms in eqn.~(\ref{eqn_djacobian_los}), and collecting terms quadratic and cubic in the gravitational potential. In this way, the second order correction $\mathcal{B}^{(2)}$ acquires terms from the perturbed geodesic as well as a lens-lens coupling term from the first order Jacobian matrix $\mathcal{A}^{(1)}$ and a term from the first order Jacobian derivative $\mathcal{B}^{(1)}$:
\begin{equation}
\mathcal{B}_{ijk}^{(2)}(\btheta,\chi) = -2\int_0^\chi\dd\chip\:(\chi-\chip)\frac{\chip^2}{\chi}
\left[
\varphi_{ijkm}x^{(1)}_m +
\varphi_{imk}\mathcal{A}^{(1)}_{mj} +
\varphi_{ijm}\mathcal{A}^{(1)}_{mk} + 
\frac{1}{\chip}
\varphi_{im}\mathcal{B}^{(1)}_{mjk}
\right].
\label{eqn_flexion_second}
\end{equation}
The third-order correction $\mathcal{B}^{(3)}$ is then given by:
\begin{eqnarray}
\mathcal{B}_{ijk}^{(3)}(\btheta,\chi) = 
& - & 2\int_0^\chi\dd\chip\:(\chi-\chip)\frac{\chip^2}{\chi}
\left[
\varphi_{ijm}\mathcal{A}^{(2)}_{mk} + 
\varphi_{imk}\mathcal{A}^{(2)}_{mj} +
\varphi_{imkn}x^{(1)}_n\mathcal{A}^{(1)}_{mj} + 
\varphi_{ijmn}x^{(1)}_n\mathcal{A}^{(1)}_{mk} +
\varphi_{imn}\mathcal{A}^{(1)}_{mj}\mathcal{A}^{(1)}_{nk} +
\frac{1}{2}\varphi_{ijkmn}x^{(1)}_mx^{(1)}_n
\right]\nonumber\\
& - &2\int_0^\chi\dd\chip\:(\chi-\chip)\frac{\chip}{\chi}
\left[
\varphi_{imn}x^{(1)}_n\mathcal{B}^{(1)}_{mjk}+\varphi_{im}\mathcal{B}^{(2)}_{mjk}
\right],
\end{eqnarray}
with the necessary subsitutions of the geodesic $x^{(1)}(\btheta,\chip)$, the Jacobians $\mathcal{A}^{(i)}(\btheta,\chip)$ and the derivatives $\mathcal{B}^{(i)}(\btheta,\chip)$, $i=1,2$. It is not straightforward to identify the terms contributing to the correction $\mathcal{B}^{(3)}$ as being due to the perturbed geodesic or due to lens-lens coupling, as different orders in perturbation theory start mixing. For brevity, we have suppressed the argument $(\geo,\chip)$ of the gravitational potential $\varphi$, and the argument $(\btheta,\chip)$ of the geodesic $\vecx^{(1)}$, the Jacobian matrices $\mathcal{A}^{(1)}$, $\mathcal{A}^{(2)}$ and of $\mathcal{B}^{(1)}$. 

We will focus on the spectra $\bra\mathcal{B}^{(1)}\mathcal{B}^{(1)}\ket$ and $\bra\mathcal{B}^{(2)}\mathcal{B}^{(2)}\ket$, and neglect the cross spectrum $\bra\mathcal{B}^{(1)}\mathcal{B}^{(2)}\ket$ which is zero for linearly evolving scales (being proportional to $\varphi^3$) and would only be nonvanishing for nonlinearly evolving scales. We focus in this work on the mode-coupling generated by lensing itself, and not that introduced by nonlinear structure formation. The amplitude of the weak lensing bispectrum relative to the spectrum is rather small on scales $\ell\lsim10^3$ \citep[see, e.g. Figs.~4 and~5 in][]{2008arXiv0803.2154S} which leads us to believe that on the scales considered here, the $\bra\mathcal{B}^{(1)}\mathcal{B}^{(2)}\ket$-term can be discarded. 

We expect that it is the Born correction which is more important for the flexion signal than the lens-lens couplings, which can be understood from physical arguments: If a light-ray is deflected from its fiducial path by $x^{(1)}_m$ it encounters changed tidal fields $\varphi_{ijkm}$ introducing flexion. The field $\varphi_{ijkm}$ varies rapidly with distance because in general the derivatives of a Gaussian random field fluctuate more rapidly when increasing order of the derivative. In contrast, lens-lens terms of the type $\varphi_{imk}\mathcal{A}^{(n)}_{mj}$ and $\varphi_{im}\mathcal{B}^{(n)}_{mjk}$ will be smaller because of the line-of-sight integration in the computation of $\mathcal{A}^{(n)}$ and $\mathcal{B}^{(n)}$, which averages out fluctuations. For these reasons, we will neglect the cross-correlation $\bra\mathcal{B}^{(1)}\mathcal{B}^{(3)}\ket$ as it contains terms of that type.

\subsection{Diagramatic representation of the perturbative expansion}
\begin{figure}
\begin{center}
\begin{tabular}{ccc}
\hline
deflection angle $\beta$ & 
Jacobian $\mathcal{A}=\partial\beta/\partial\theta$ & 
derivative $\mathcal{B}=\partial\mathcal{A}/\partial\theta$\\
\hline
\resizebox{4cm}{!}{\includegraphics{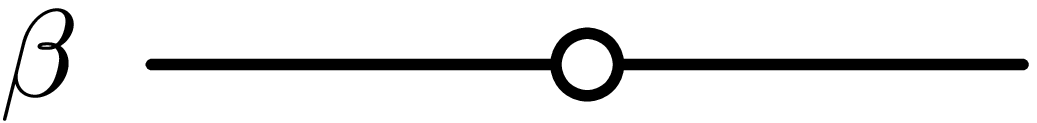}} & 
\resizebox{4cm}{!}{\includegraphics{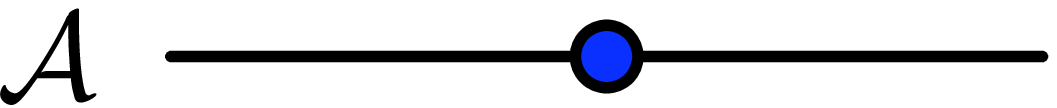}} &
\resizebox{4cm}{!}{\includegraphics{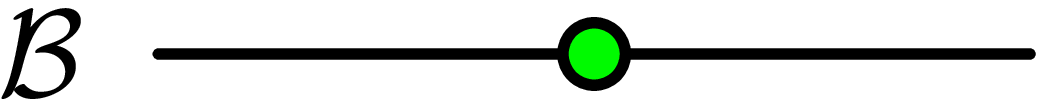}} \\
\hline
&
\resizebox{4cm}{!}{\includegraphics{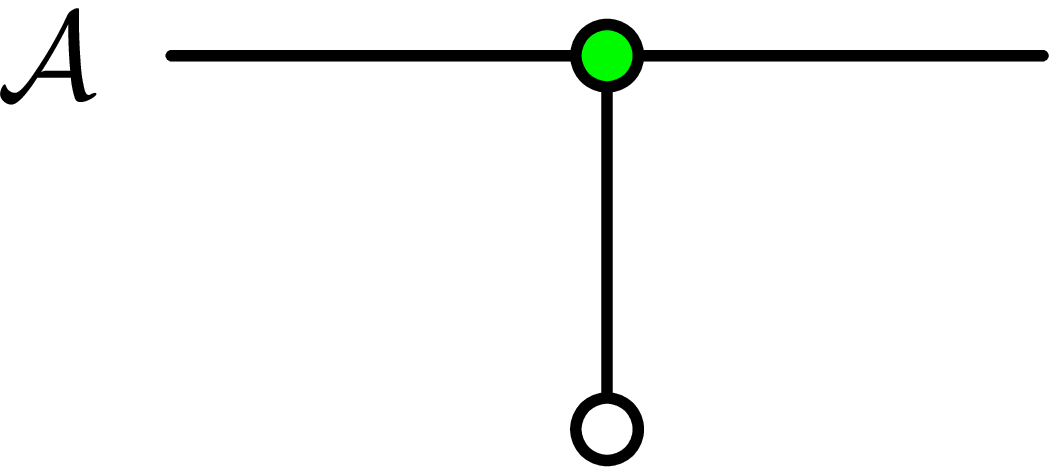}} &
\resizebox{4cm}{!}{\includegraphics{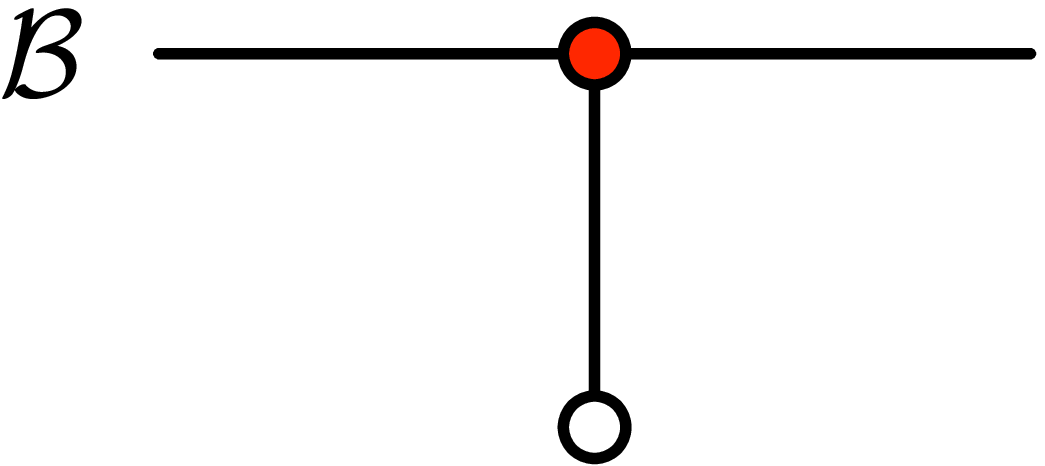}} \\
\hline
&
\resizebox{4cm}{!}{\includegraphics{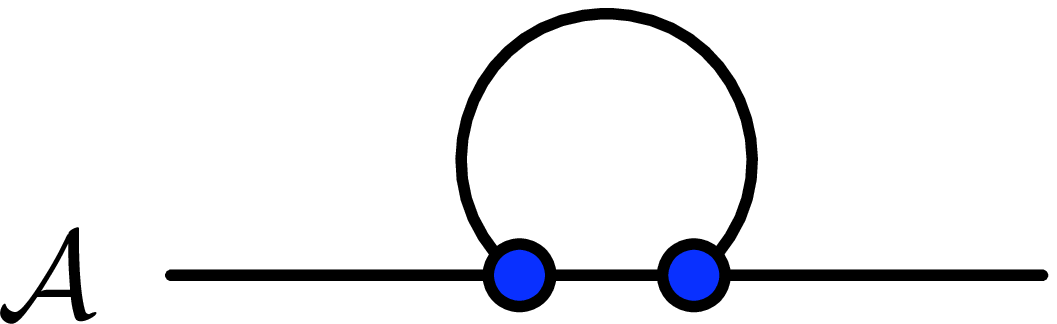}} &
\resizebox{4cm}{!}{\includegraphics{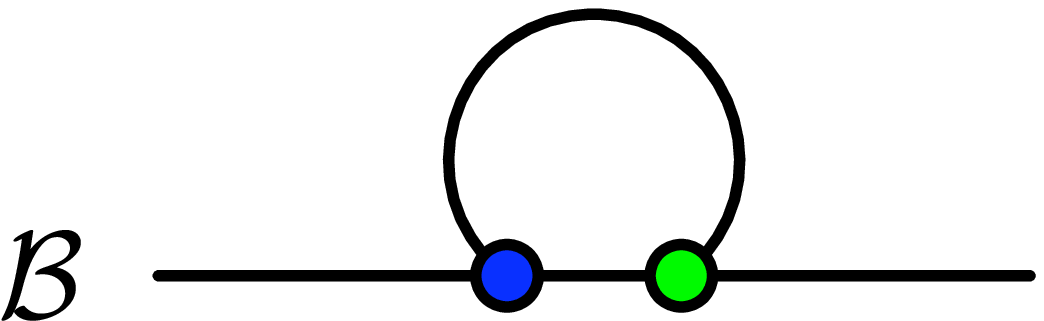}} \\
&
&
\resizebox{4cm}{!}{\includegraphics{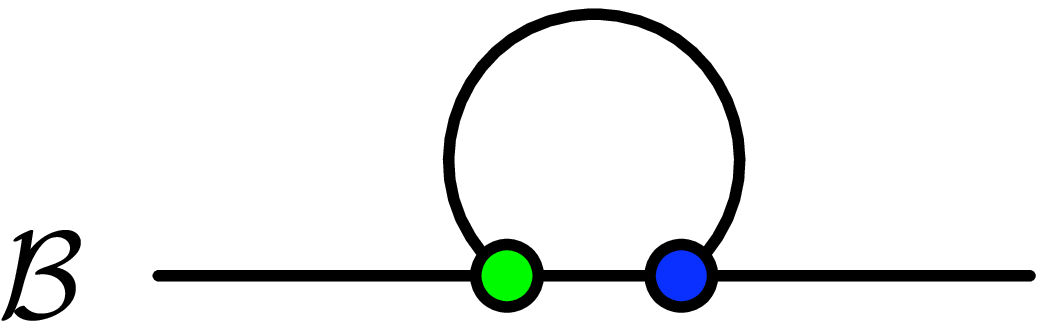}} \\
\hline
\end{tabular}
\end{center}
\caption{Diagramatic representation of the correction terms in the line of sight expressions for the deflection angle $\beta$, the Jacobian $\mathcal{A}$ and the derivative $\mathcal{B}$. The shading of the vertices indicates the order of the derivative of the potential: first order (white), second order (blue), third order (green), fourth order (red). The first row assumes that all approximations are in place, the second row describes the Born-corrections, and the third row sums up the lens-lens couplings, all of which are non-local. The number of vertices is equal to the order of perturbation, and therefore to the power of the gravitational potential.}
\label{fig_comic}
\end{figure}

A visual representation of the perturbative corrections to the lensing effects is given in Fig.~\ref{fig_comic}. This shows graphs corresponding to the corrections to the lensing deflection angle $\beta$ (first column), the Jacobian $\mathcal{A}$ (centre column) and the derivative $\mathcal{B}$ (right column) for a simplified approximative calculation (first row), for the corrections to Born (second row) and the lens-lens couplings (third row). With all approximations in place (first row), the lensing deflection angle $\beta$ contains gradients of the gravitational potential, the Jacobian $\mathcal{A}$ contains second derivatives and the derivative $\mathcal{B}$ of the Jacobian contains third derivatives of the gravitational potential, in accordance with Fig.~\ref{fig_flexion}. Under these approximations, the interaction of the light ray with a gravitational potential is local, and the interchangability of partial derivatives translates to the symmetry under index exchange of $\mathcal{A}$ and $\mathcal{B}$.

If first-order Born-corrections to the geodesic are taken into account (second row), the Jacobian $\mathcal{A}$ is corrected by a term linking the third derivative on the fiducial ray with the displacement from the fiducial geodesic (c.f. centre panel in the second row, corresponding to the second term in eqn.~\ref{eqn_jacobian_second}), and the derivative $\mathcal{B}$ measures the contraction of the fourth derivative of the potential along the fiducial ray with the displacement (right panel of the second row, depicting the first term of eqn.~\ref{eqn_flexion_second}). 

Lens-lens coupling (third row), i.e. the interplay between multiple distortions along the line-of-sight, appears in the Jacobian $\mathcal{A}$ as a contraction between two second derivatives of the gravitational potential (centre panel in the third row of Fig.~\ref{fig_comic}, first term in eqn.~\ref{eqn_jacobian_second}). The equivalent corrections for the flexion signal consist of contractions of second and third derivatives of the gravitational potential (right panel in the third row): firstly, there is a contraction of the third derivative of the gravitational potential with the integrated second derivative, i.e. the Jacobian $\mathcal{A}$ (second and third term in eqn.~\ref{eqn_flexion_second}) and secondly, a contraction of the second derivative with the integrated third derivative $\mathcal{B}$ (fourth term in eqn.~\ref{eqn_flexion_second}), both of these taking place along the unperturbed ray path.  It is important that in this panel both graphs with interchanged vertices appear, which is a consequence of the reversability of the light path in geometrical optics. If only one of the graphs was present, the reversability would be violated because a specific direction of propagation would be singled out. Due to the fact that both graphs with interchanged vertex order are present, no specific direction of propagation is preferred over the other.

So far, we have considered corrections to second order involving products of potentials, which corresponds to the fact that there are two vertices in each graph, but with these graphical rules, it is quite easy to generalise the expressions for $\mathcal{A}$ and $\mathcal{B}$ (or even $\partial\mathcal{B}/\partial\theta$) to higher orders bearing in mind that we only acquire corrections of order $\varphi^2$ and therefore would follow the pattern outlined in Fig.~\ref{fig_comic}.

Quite generally, the lowest order quantities are symmetric under index exchange, reflecting the interchangeability of partial derivatives of the projected gravitational potential, whereas higher-order corrections violate this feature. Hence the Jacobian $\mathcal{A}^{(1)}$ is symmetric, with $\mathcal{A}^{(2)}$ breaking this symmetry. In complete analogy, $\mathcal{B}^{(1)}$ is symmetric under index exchange, where again the higher-order corrections $\mathcal{B}^{(2)}$ and $\mathcal{B}^{(3)}$ introduce a symmetry-breaking in the first two indices. Conversely, a useful consistency check is the fact that the expressions for $\mathcal{B}^{(2)}$ and $\mathcal{B}^{(3)}$ derived above respect the symmetry of $\mathcal{B}$ under exchange of the last two indices, in accordance with the definition as $\mathcal{B}_{ijk}\equiv\partial\mathcal{A}_{ij}/\partial\theta_k=\partial\mathcal{A}_{ik}/\partial\theta_j=\mathcal{B}_{ikj}$. As a last point we mention that the flexion signal has more terms contributing to corrections due to dropping the Born-approximation and lens-lens coupling, compared to the Jacobian. At first order in perturbation theory, there are twice as many terms contributing relative to shear, and at second order there are four times as many terms relative to shear. We need to examine in later sections whether this increase in number of terms corresponds to an increase in net importance of the corrections at this order.

\subsection{Decomposition in Pauli- and Dirac-matrices}
At this stage, in order to make progress in calculating the corrections, it will be very useful to introduce a decomposition of $\mathcal{A}$ and $\mathcal{B}$ in terms of matrices. The Jacobian matrix $\mathcal{A}$ can be decomposed with the Pauli-matrices $\sigma_\alpha$, because they constitute a basis for the vector space of $2\times2$ matrices,
\begin{equation}
\psi \equiv \mathrm{id}(2) - \mathcal{A} \simeq -\mathcal{A}^{(1)} - \mathcal{A}^{(2)}
= \sum_{\alpha=0}^3\: a_\alpha\sigma_\alpha 
= \kappa\sigma_0 + \gamma_+\sigma_1 - \ci\rho\sigma_2 + \gamma_\times\sigma_3,
\end{equation}
\citep[c.f.][]{1972hmf..book.....A, 2005mmp..book.....A}, where $\mathrm{id}(n)$ denotes the $n$-dimensional unit matrix. The Pauli-matrices are defined as
\begin{equation}
\sigma_0 = \left(
\begin{array}{cc}
+1 & 0 \\ 0 & +1
\end{array}
\right),\quad
\sigma_1 = \left(
\begin{array}{cc}
0 & +1 \\ +1 & 0
\end{array}
\right),\quad
\sigma_2 = \left(
\begin{array}{cc}
0 & -\ci \\ +\ci & 0
\end{array}
\right),\quad
\sigma_3 = \left(
\begin{array}{cc}
+1 & 0 \\ 0 & -1
\end{array}
\right),
\end{equation}
with the properties $\sigma_\alpha^2=\mathrm{id}(2)$ and $\trace(\sigma_\alpha)=0$ for $\alpha=1,2,3$. Due to the property $\sigma_\alpha\sigma_\beta = \mathrm{id}(2)\delta_{\alpha\beta} + \ci\epsilon_{\alpha\beta\gamma}\sigma_\gamma$ of the Pauli-matrices, the coefficients $a_\alpha$ can be recovered by using 
$a_\alpha = \frac{1}{2}\trace(\psi\sigma_\alpha)$.

 The coefficients in the decomposition can be identified as the convergence $\kappa=a_0$, the two components of shear $\gamma_+=a_1$, $\gamma_\times=a_3$ and the image rotation $\rho=a_2$. Due to the symmetry of the first order Jacobian matrix $\mathcal{A}^{(1)}$, gravitational lensing is only able to excite three of the four possible image distortions. To first order, it is only possible to observe convergence and the two components of shear. If the Jacobian $\mathcal{A}$ becomes non-symmetric because of the contribution $\mathcal{A}^{(2)}$, then the $\rho\sigma_2$ term becomes permissible and the image distortion can include a rotation \citep[][ for an application]{2000ApJ...530..547J, 2009MNRAS.396.2167B}.

In order to carry out an analogous decomposition of the derivative $\mathcal{B}$ of the Jacobian, which is a $2\times2\times2$-tensor, we recast it into a $4\times4$ symmetric block-diagonal matrix, $\mathcal{B}_{j+2i,k+2i}=\mathcal{B}_{ijk}$,
\begin{equation}
\mathcal{B}_{jk} \equiv \left(
\begin{array}{cc}
\mathcal{B}_{0jk} & 0\\ 
0& \mathcal{B}_{1jk}
\end{array}
\right),
\label{eqn_remapping}
\end{equation}
where we use the same symbol $\mathcal{B}$ as confusion is unlikely. Matrix~(\ref{eqn_remapping}) evidently shows the six flexion degrees of freedom, due to the symmetry of $\mathcal{B}_{jk}$. $\mathcal{B}_{jk}$ can be decomposed with the Dirac-matrices $\Delta_\gamma$:
\begin{equation}
\mathcal{B} = \sum_{\gamma=0}^5\: b_\gamma \Delta_\gamma
\quad\mathrm{with}\quad
b_\gamma = \frac{1}{4}\mathrm{tr}(\mathcal{B}\Delta_\gamma).
\end{equation}
The Dirac-matrices $\Delta_\gamma$ are a generalisation of the Pauli-matrices $\sigma_\beta$ \citep{2005mmp..book.....A}, of which we select the symmetric subset
\begin{equation}
\Delta_0 = \left(
\begin{array}{cc}
+\sigma_0 & \\
& +\sigma_0
\end{array}
\right),\quad
\Delta_1 = \left(
\begin{array}{cc}
+\sigma_1 &\\
& +\sigma_1 
\end{array}
\right),\quad
\Delta_2 = \left(
\begin{array}{cc}
+\sigma_3 & \\
& +\sigma_3
\end{array}
\right),
\end{equation}

\begin{equation}
\Delta_3 = \left(
\begin{array}{cc}
+\sigma_0 & \\
& -\sigma_0
\end{array}
\right),\quad
\Delta_4 = \left(
\begin{array}{cc}
+\sigma_1 &\\
& -\sigma_1 
\end{array}
\right),\quad
\Delta_5 = \left(
\begin{array}{cc}
+\sigma_3 & \\
& -\sigma_3
\end{array}
\right),
\end{equation}
with the unit element $\Delta_0\equiv \mathrm{id}(4)$. These matrices have the properties $\Delta_\gamma ^2=\mathrm{id}(4)$, $\trace(\Delta_\gamma)=0$ for $\gamma\geq 1$ and form a basis set of the vector space of block-diagonal symmetric $4\times 4$-matrices, because the $\sigma_\beta$, $\beta=0,1,3$ are symmetric, $\sigma_\beta^t=\sigma_\beta$ . The fact that there are exactly these six $\Delta_\gamma$-matrices is consistent with the six degrees of freedom of the tensor $\mathcal{B}_{ijk}$, as explained earlier. The selected subset of Dirac matrices are a closed group under multiplication, and the product $\Delta_\alpha\Delta_\beta$ of two different $\Delta_\beta$-matrices is always traceless such that the decomposition $b_\gamma=\frac{1}{4}\trace(\mathcal{B}\Delta_\gamma)$ is always possible.

\subsection{Rotations and coordinate-independent representation}
The transformation properties of the $\mathcal{B}_{ij}$-tensor and hence of the coefficients $b_\beta$ under coordinate rotations are complicated, which is due to the transformation law $\mathcal{B}_{ijk} = \sum_{i^\prime}\sum_{j^\prime}\sum_{k^\prime}R_{ii^\prime}R_{jj^\prime}R_{kk^\prime}\mathcal{B}_{i^\prime j^\prime k^\prime}$ with three $2\times2$ rotation matrices $R_{ii^\prime}$. This is in contrast to the decomposition of the Jacobian $\mathcal{A}$ into Pauli-matrices, where the coefficient $\kappa$ is invariant under orthogonal transformations of the coordinate system, and the two components of shear transform into each other. They can be combined to form the complex shear $\gamma=\gamma_++\ci\gamma_\times$ with the transformation property $\gamma\rightarrow\gamma\exp(2\ci\varphi)$ under rotations by $\varphi$. The complexity arises when the tensor $\mathcal{B}_{ijk}$ is mapped onto the block diagonal matrix $\mathcal{B}_{jk}$ and the transformations of the individual indices get mixed. The decomposition of $\mathcal{B}$ with $\Delta_0$ and $\Delta_3$ gives rise to a quantity which transforms like a vector, i.e. maps onto itself after rotation by $2\pi$, and which is non-zero when the Born approximation is included (see Sect.~4.4).  We recognise this as the vectorial spin-1 flexion $\mathcal{F}$; thus $\mathcal{F}$ can be recovered by decomposition of $\mathcal{B}$ with $\Delta_0$ and $\Delta_3$. Decomposition of $\mathcal{B}$ with the remaining $\Delta$-matrices has a more complex transformation property; however under the Born approximation the decompositions with $\Delta_2$ and $\Delta_5$ are non-zero, so these are associatied with the flexion $\mathcal{G}$. The decompositions with $\Delta_1$ and $\Delta_4$ are unactivated in straight light paths, and correspond to the twist components.

\section{Angular spectra}\label{sect_spectra}

\subsection{Angular power spectra}
We now proceed to calculate angular power spectra for the various flexion components, including the perturbation corrections we have described above. To lowest order in perturbation theory, the two-point correlation function of the Jacobian matrices $\mathcal{A}$ and $\mathcal{B}$ and hence the angular spectra are determined by a projection of the power spectrum of the gravitational potential $P_{\varphi}(k)$. As shown in Sect.~\ref{sect_lensing}, dropping the Born-approximation introduces corrections of order $\mathcal{O}(\varphi^2)$; these real-space products of $\phi$ become convolutions in Fourier-space, so in this section we will find terms containing the squared spectrum $P_{\varphi}^2(k)$ with a mode-coupling kernel. We will truncate the perturbative expansion and discard terms of order $\varphi^3$ because we expect them to be small compared to the corrections $\mathcal{O}(\varphi^2)$, and due to reasons of practicality: in the power spectra, the corrections $\mathcal{O}(\varphi^2)$ give rise to $6$ terms, whereas the corrections $\mathcal{O}(\varphi^3)$ cause a total of $8+{8\choose 3}=64$ terms.

We calculate the correlation function of $a_\alpha$, which are the components of the Jacobian $\mathcal{A}$ via
\begin{equation}
\bra a_\alpha(\vecl) a_\beta^*(\veclp)\ket = 
\frac{1}{2^2}\bra\trace(\mathcal{A}(\vecl)\sigma_\alpha)\:\trace(\mathcal{A}(\veclp)\sigma_\beta)^*\ket = 
\frac{1}{2^2}\sum_{a,b}\sum_{i,j}\:\bra\mathcal{A}_{ab}(\vecl)\mathcal{A}^*_{ij}(\veclp)\ket\:
\sigma_{\alpha,ba}^{\vphantom{*}}\sigma_{\beta,ji}^*,
\label{eqn_a_2point_var}
\end{equation}
and apply an analogous relation for the correlation function of the flexion components $b_\gamma$,
\begin{equation}
\bra b_\gamma(\vecl) b_\delta^*(\veclp)\ket =
\frac{1}{4^2}\bra\trace(\mathcal{B}(\vecl)\Delta_\gamma)\:\trace(\mathcal{B}(\veclp)\Delta_\delta)^*\ket = 
\frac{1}{4^2}\sum_{a,b}\sum_{i,j}\:\bra\mathcal{B}_{ab}(\vecl)\mathcal{B}^*_{ij}(\veclp)\ket\:
\Delta_{\gamma,ba}^{\vphantom{*}}\Delta_{\delta,ji}^*.
\label{eqn_b_2point_var}
\end{equation}
In total, the expression eqn.~(\ref{eqn_a_2point_var}) contains 4 terms, and eqn.~(\ref{eqn_b_2point_var}) contains 16 terms, where the signs are given by contraction with $\sigma_{\alpha,ki}^{\vphantom{*}}\sigma_{\beta,ba}^*$ and $\Delta_{\gamma,ba}^{\vphantom{*}}\Delta_{\delta,ji}^*$, respectively. We introduce the angular power spectra $C_{abij}^{\mathcal{A}}(\ell)$ of the Jacobian $\mathcal{A}$,
\begin{equation}
\bra\mathcal{A}_{ab}(\vecl)\mathcal{A}_{ij}^*(\veclp)\ket = 
(2\pi)^2\dirac(\vecl-\veclp) C_{abij}^{\mathcal{A}}(\ell),
\label{eqn_shear_contraction}
\end{equation}
and $C_{abij}^{\mathcal{B}}(\ell)$ of the derivative $\mathcal{B}$ of the Jacobian,
\begin{equation}
\bra\mathcal{B}_{ab}(\vecl)\mathcal{B}_{ij}^*(\veclp)\ket = 
(2\pi)^2\dirac(\vecl-\veclp) C_{abij}^{\mathcal{B}}(\ell),
\end{equation}
such that the flexion and twist spectra $C^\mathcal{B}_{\gamma\delta}(\ell)$, $\gamma,\delta=0\ldots5$, can be defined with ($\Delta^+=\Delta$, i.e. $\Delta^t=\Delta$ and $\Delta^*=\Delta$ for the selected subset):
\begin{equation}
C^\mathcal{B}_{\gamma\delta}(\ell) = 
\frac{1}{4^2}\sum_{a,b}\sum_{i,j}C_{abij}^\mathcal{B}(\ell)\:\Delta_{\gamma,ba}^{\vphantom{*}}\Delta_{\delta,ji}^* =
\frac{1}{4^2}\sum_{a,b}\sum_{i,j}C_{abij}^\mathcal{B}(\ell)\:\Delta_{\gamma,ab}\Delta_{\delta,ij},
\label{eqn_flexion_contraction}
\end{equation}
in analogy to the four spectra $C^\mathcal{A}_{\alpha\beta}(\ell)$, $\alpha,\beta=0\ldots3$, describing convergence ($\alpha=0$), the two components of shear ($\alpha=1,3$) and the image rotation ($\alpha=2$),
\begin{equation}
C^\mathcal{A}_{\alpha\beta}(\ell) = 
\frac{1}{2^2}\sum_{a,b}\sum_{i,j}C_{abij}^\mathcal{A}(\ell)\:\sigma_{\alpha,ba}^{\vphantom{*}}\sigma_{\beta,ji}^*.
\end{equation}
The power spectrum $P_{\varphi}(k,a)$ of the potential $\varphi\equiv\Phi/c^2$ at the epoch $a$ follows from the comoving Poisson equation $\Delta\varphi=3H_0^2\Omega_m/(2a)\delta$ and is related to the density power spectrum $P_{\delta}(k,a)$ by
\begin{equation}
P_{\varphi}(k,a) = \frac{9\Omega_m^2}{4}\left(\frac{D_+(a)}{a}\right)^2\frac{P_{\delta}(k)}{(\chi_Hk)^4},
\end{equation}
where the Hubble distance $\chi_H=c/H_0$ makes the $k^{-4}$-factor dimensionless. For simplicity, we only consider linearly growing scales in this study, but will make an argument to extrapolate to smaller scales in Sect.~\ref{sect_corrected_spectra}.

\subsection{Born-approximation}
In order to project the source term power spectrum $P_\delta(k)$ in order to estimate the angular power spectrum $C_\kappa(\ell)$ of the observable, the flat-sky Limber-equation is used \citep{1954ApJ...119..655L},
\begin{equation}
\kappa=\int_0^{\chi_s}\dd\chi\: W(\chi)\:\delta(\chi)
\longrightarrow
C_\kappa(\ell)=\int_0^{\chi_s}\frac{\dd\chi}{\chi^2}\: W(\chi)^2P_\delta(k=\ell/\chi),
\label{eqn_lensing_spectrum}
\end{equation}
with the lensing efficiency function $W(\chi)$ for sources placed at the comoving distance $\chi_s$,
\begin{equation}
W(\chi) = \frac{3H_0^2\Omega_m}{2c^2}\frac{\chi_s-\chi}{\chi_s}\frac{D_+(\chi)}{a}\chi.
\end{equation}
The integration in eqn.~(\ref{eqn_lensing_spectrum}) is extended to the comoving distance of the lensed background galaxies $\chi_s$. The source term spectrum of the shear-related image distortions is given by four-fold differentiation of the potential power spectrum, which results in the angular power spectrum $C_{abij}^\mathcal{A}(\ell)$ after Limber-projection,
\begin{equation}
C_{abij}^\mathcal{A}(\vecl) = 2^2\int_0^{\chi_s}\frac{\dd\chi}{\chi^4}\:\left(\frac{\chi_s-\chi}{\chi_s}\right)^2\ell_a\ell_b\:\ell_i\ell_j\: P_{\varphi}(k=\ell/\chi,\chi).
\end{equation}
Spectra of the individual lensing observables can then be obtained by contraction with $\sigma_{\beta,ki}^{\vphantom{*}}\sigma_{\beta,jl}^*$ (c.f. eqn.~\ref{eqn_shear_contraction}). This technique is very useful for the far larger number of correction terms in the flexion specta. The relevant flexion spectra are derived in complete analogy: the source term spectrum follows from six-fold differentiation of the power spectrum of the gravitational potential, with the corresponding implicit projection onto $C_{bcjk}^\mathcal{B}(\ell)$,
\begin{equation}
C_{b+2a,c+2a,j+2i,k+2i}^\mathcal{B}(\vecl) = 2^2\int_0^{\chi_s}\frac{\dd\chi}{\chi^4}\:\left(\frac{\chi_s-\chi}{\chi_s}\right)^2\ell_a\ell_b\ell_c\:\ell_i\ell_j\ell_k\: P_{\varphi}(k=\ell/\chi,\chi).
\end{equation}
Again, the spectra of the flexion components follow from contraction with $\Delta_{\gamma,ab}\Delta_{\delta,ij}$, according to eqn.~(\ref{eqn_flexion_contraction}), after recasting the rank-3 tensor $\mathcal{B}_{ijk}$ into a $4\times4$-matrix $\mathcal{B}_{jk}$ using eqn.~(\ref{eqn_remapping}).

Due to the transformation properties of $\mathcal{B}$, we do not attempt to define a coordinate-independent representation equivalent to the $E$- and $B$-modes of weak shear \citep{1996astro.ph..9149S, 1998MNRAS.301.1064K, 2001ApJ...554...67H}. Instead, we provide spectra of the expansion coefficients $b_\gamma$ by choosing the coordinate frame such that the separation vector $\vecl$ coincides with the $\ell_x$-axis of the Fourier-frame. In this frame, only the component $C^\mathcal{B}_{0000}(\ell)$ is nonzero, which leads to nonzero projections with $\Delta_0$, $\Delta_2$, $\Delta_3$ and $\Delta_5$. The vanishing projections are those with matrices constructed from $\sigma_1$, which is sensitive to off-diagonal elements in $\mathcal{B}_{ab}$, all of which are zero. This corresponds to the absence of a twist mode under the Born-approximation. In contrast, matrices constructed from $\sigma_0$ and $\sigma_3$ measure differences between the diagonal elements inside each block, as well as between the blocks, and are therefore non-vanishing.

\subsection{Corrections due to the perturbed geodesic, lens-lens coupling and lens-flexion coupling}
In this section, we compute the spectrum $\bra\mathcal{B}^{(2)}\mathcal{B}^{(2)}\ket$ of the second order corrections $\mathcal{B}^{(2)}$ which are quadratic in the gravitational potential. As explained in Sect.~\ref{sect_optical}, we expect the terms in $\bra\mathcal{B}^{(2)}\mathcal{B}^{(2)}\ket$ to be the dominating contributions to the flexion spectrum. The correction $\mathcal{B}^{(2)}_{ijk}$ to the Jacobian derivative, at which we truncate the perturbative expansion, will comprise one Born-term, two lens-terms and one flexion-lens term and reads
\begin{equation}
\mathcal{B}^{(2)}_{ijk}(\btheta,\chi) = 
4
\int_0^{\chi}\dd\chip\:(\chi-\chip)\frac{\chip^2}{\chi}
\int_0^{\chip}\dd\chipp\:(\chip-\chipp)
\Bigl[
\underbrace{\vphantom{\left(\frac{\chipp}{\chip}\right)^2}\:\varphi_{ijkm}\varphi_{m}}_{\equiv S^{(b)}_{ijk}} +
\underbrace{\vphantom{\left(\frac{\chipp}{\chip}\right)^2}\left(\frac{\chipp}{\chip}\right)
\left(\varphi_{imk}\varphi_{mj} + \varphi_{ijm}\varphi_{mk}\right)}_{\equiv S^{(l)}_{ijk}} +
\underbrace{\left(\frac{\chipp}{\chip}\right)^2\varphi_{im}\varphi_{mjk}}_{\equiv S^{(f)}_{ijk}}
\Bigl],
\end{equation}
after substitution of the first-order expressions for $x^{(1)}_i$, $\mathcal{A}^{(1)}_{ij}$ and $\mathcal{B}^{(1)}_{ijk}$ into eqn.~(\ref{eqn_flexion_second}). Again, the symmetry $\mathcal{B}_{ijk}=\mathcal{B}_{ikj}$ remains conserved at all orders of perturbation theory. In particular the two lens-lens terms transform into each other under exchange $j\leftrightarrow k$ and are both necessary to conserve this particular symmetry. 

\begin{figure}
\begin{center}
\resizebox{0.5\hsize}{!}{\includegraphics{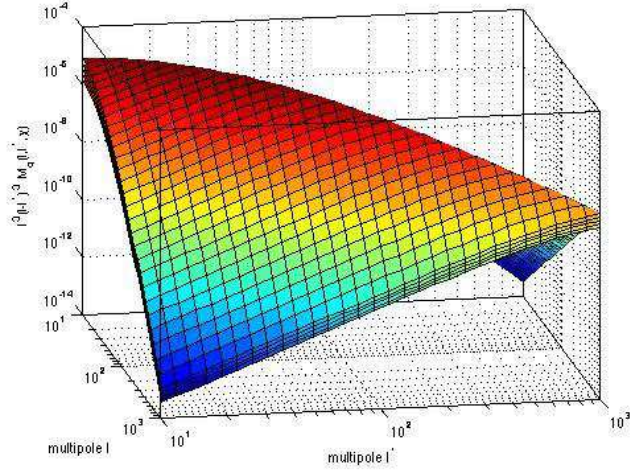}}
\end{center}
\caption{Mode coupling functions $M_q(\ell,\ell^\prime,\chi)$, for $q=0\ldots4$ (from top to bottom) used in the corrections to the angular flexion spectra, computed for a distance of $\chi=1~\mathrm{Gpc}/h$, which corresponds to a redshift of $z\simeq0.36$ in $\Lambda$CDM with $\Omega_m=0.25$. The angle $\phi_{\lprime}$ between $\vecl$ and $\veclp$ is fixed at a value of $\pi/2$.}
\label{fig_coupling}
\end{figure}

Consequently, the power spectrum of $\mathcal{B}$ acquires 6 correctional terms, 3 auto-spectra and 3 cross-spectra. The dependence of the terms in $\mathcal{B}$ on the square of the gravitational potential maps onto convolutions of the potential in Fourier space, together with contraction of the free indices. The $\dd\chi^\prime$-integration will give rise to mode-couplings, weighted by different powers of $\chi^\prime/\chi$ in each of the source terms $S_{ijk}(\vecl,\chi)$:
\begin{eqnarray}
S_{ijk}^{(b)}(\vecl,\chi) & = & 
\int\frac{\dd^2\lprime}{(2\pi)^2}\:
\int_0^\chi\dd\chip\:(\chi-\chip)\:\varphi_{ijkm}(\veclp)\varphi_m(\vecl-\veclp),\\
S_{ijk}^{(l)}(\vecl,\chi) & = & 
\int\frac{\dd^2\lprime}{(2\pi)^2}\:
\int_0^\chi\dd\chip\:(\chi-\chip)\left(\frac{\chip}{\chi}\right)
\left(\varphi_{imk}(\veclp)\varphi_{mj}(\vecl-\veclp) + \varphi_{ijm}(\veclp)\varphi_{mk}(\vecl-\veclp)\right),\\
S_{ijk}^{(f)}(\vecl,\chi) & = &
\int\frac{\dd^2\lprime}{(2\pi)^2}\:
\int_0^\chi\dd\chip\:(\chi-\chip)\left(\frac{\chip}{\chi}\right)^2\varphi_{im}(\veclp)\varphi_{mjk}(\vecl-\veclp).
\end{eqnarray}
These source terms can be combined to yield the spectrum:
\begin{equation}
\bra S_{abc}(\vecl,\chi)S_{ijk}^*(\veclp,\chip)\ket = 
\frac{(2\pi)^2}{\chi^6}\dirac(\chi-\chip)\dirac(\vecl-\veclp) P_{abcijk}(\vecl,\chi),
\end{equation}
which can then be projected for obtaining the angular spectra:
\begin{equation}
C_{b+2a,c+2a,j+2i,k+2i}(\vecl) = 
4^2\int_0^{\chi_s}\frac{\dd\chi}{\chi^4}\:
\left(\frac{\chi_s-\chi}{\chi_s}\right)^2 P_{abcijk}(\vecl,\chi).
\end{equation}
The resulting power spectra between the Born-, lens- and flexion-terms are obtained by applying the Wick theorem \citep[c.f.][]{2008cmbg.book.....D} to the 4-point correlation of the gravitational potential. It is worth noting that the various resulting spectra have very similar forms, differing only in the extent of mode coupling and dependence on $\ell$:
\begin{eqnarray}
P_{abcijk}^{(bb)}(\vecl,\chi) & = & 
\int\frac{\dd^2\lprime}{(2\pi)^2}\:\left[\veclp(\vecl-\veclp)\right]^2 \:
\left(\lprime_a\lprime_b\lprime_c\:\:\lprime_i\lprime_j\lprime_k\right)
\:M_0(\vecl,\veclp,\chi),\\ 
P_{abcijk}^{(ll)}(\vecl,\chi)  & = & 
\int\frac{\dd^2\lprime}{(2\pi)^2}\:\left[\veclp(\vecl-\veclp)\right]^2 \:
\left(
\lprime_a\dlp_b\lprime_c\lprime_i\dlp_j\lprime_k +
\lprime_a\dlp_b\lprime_c\lprime_i\lprime_j\dlp_k +
\right. \nonumber\\ && 
\hphantom{\int\frac{\dd^2\lprime}{(2\pi)^2}\:\left[\veclp(\vecl-\veclp)\right]^2 \:\:}\left.
\lprime_a\lprime_b\dlp_c\lprime_i\dlp_j\lprime_k +
\lprime_a\lprime_b\dlp_c\lprime_i\lprime_j\dlp_k
\right)
\:M_2(\vecl, \veclp,\chi),\\ 
P_{abcijk}^{(ff)}(\vecl,\chi)  & = & 
\int\frac{\dd^2\lprime}{(2\pi)^2}\:\left[\veclp(\vecl-\veclp)\right]^2 \:
\left(\lprime_a\dlp_b\dlp_c\:\lprime_i\dlp_j\dlp_k\right)
\:M_4(\vecl, \veclp,\chi),
\end{eqnarray}
The corresponding cross spectra, where a symmetrisation has been applied, are given by
\begin{eqnarray}
P_{abcijk}^{(bl)}(\vecl,\chi) & = & 
\int\frac{\dd^2\lprime}{(2\pi)^2}\:\left[\veclp(\vecl-\veclp)\right]^2 \:
\left(
\lprime_a\dlp_b\lprime_c\lprime_i\lprime_j\lprime_k +
\lprime_a\lprime_b\dlp_c\lprime_i\lprime_j\lprime_k +
\right. \nonumber\\ && 
\hphantom{\int\frac{\dd^2\lprime}{(2\pi)^2}\:\left[\veclp(\vecl-\veclp)\right]^2 \:\:}\left.
\lprime_a\lprime_b\lprime_c\lprime_i\dlp_j\lprime_k +
\lprime_a\lprime_b\lprime_c\lprime_i\lprime_j\dlp_k
\right)
\:M_1(\vecl, \veclp,\chi),\\ 
P_{abcijk}^{(lf)}(\vecl,\chi) & = & 
\int\frac{\dd^2\lprime}{(2\pi)^2}\:\left[\veclp(\vecl-\veclp)\right]^2 \:
\left(
\lprime_a\dlp_b\lprime_c\lprime_i\dlp_j\dlp_k +
\lprime_a\lprime_b\dlp_c\lprime_i\dlp_j\dlp_k +
\right. \nonumber\\ && 
\hphantom{\int\frac{\dd^2\lprime}{(2\pi)^2}\:\left[\veclp(\vecl-\veclp)\right]^2 \:\:}\left.
\lprime_a\dlp_b\dlp_c\lprime_i\dlp_j\lprime_k +
\lprime_a\dlp_b\dlp_c\lprime_i\lprime_j\dlp_k
\right)
\:M_3(\vecl, \veclp,\chi),\\ 
P_{abcijk}^{(bf)}(\vecl,\chi) & = & 
\int\frac{\dd^2\lprime}{(2\pi)^2}\:\left[\veclp(\vecl-\veclp)\right]^2 \:
\left(
\lprime_a\lprime_b\lprime_c\:\lprime_i\dlp_j\dlp_k +
\lprime_a\dlp_b\dlp_c\:\lprime_i\lprime_j\lprime_k
\right)
\:M_2(\vecl, \veclp,\chi). 
\end{eqnarray}
Common to all spectra is the mode coupling function $M_q$, which arises from the fact that the corrections are proportional to the square of the gravitational potential in real space, which translates into a convolution in Fourier-space. The $\dd^2\lprime$-integration is carried out in polar coordinates, $\dd^2\lprime=\lprime\dd\lprime\dd\phi_{\lprime}$ with $\vecl$ aligned along the $\ell_x$-axis such that $\vecl=\ell(1,0)$ and $\veclp=\lprime(\cos\phi_{\lprime},\sin\phi_{\lprime})$. The mode coupling $M_q(\vecl,\veclp,\chi)$ reads:
\begin{equation}
M_q(\vecl,\veclp,\chi) = 2\int_0^\chi\frac{\dd\chip}{\chip^6}\:(\chi-\chip)^2\left(\frac{\chip}{\chi}\right)^q\:P_\varphi(\left|\veclp\right|,\chip)P_\varphi(\left|\vecl-\veclp\right|,\chip).
\end{equation}
The mode coupling kernels $M_q(\ell,\lprime,\chi)$ are depicted for a fixed distance $\chi=1~\mathrm{Gpc}/h$ and a fixed angle $\phi_{\lprime}$ between $\vecl$ and $\veclp$ of $\pi/2$ in Fig.~\ref{fig_coupling}, for the range of indices $q=0\ldots4$ considered here. The strongest mode coupling is between adjacent $\ell$-modes and drops rapidly with increasing distance in Fourier-space. Fig.~\ref{fig_coupling} confirms our qualitative argumentation at the end of Sect.~\ref{sect_optical} concerning the relative magnitude of the Born- and lens-lens terms: $M_q(\ell,\lprime,\chi)$ assumes the largest values for small $q$, and in particular for $q=0$, which is the spectrum of the Born-correction only, and decreases if lens-lens terms are considered instead. This provides a quantitative argument for our expectation that the largest contributions originate from the Born-terms in the $\bra\mathcal{B}^{(2)}\mathcal{B}^{(2)}\ket$-spectrum.

\subsection{Corrected angular flexion spectra}\label{sect_corrected_spectra}
\begin{figure}
\begin{center}
\resizebox{0.5\hsize}{!}{\includegraphics{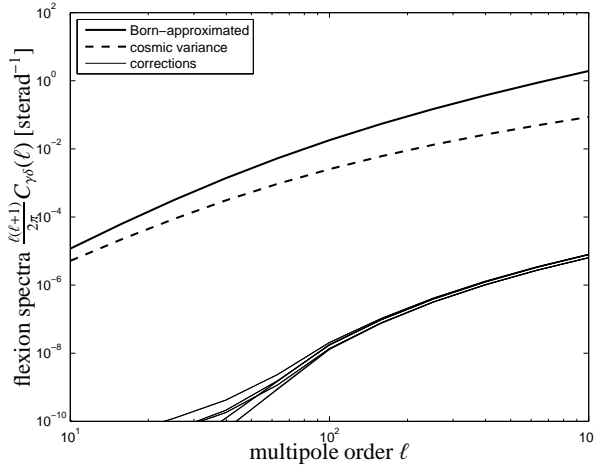}}
\end{center}
\caption{Flexion angular power spectra $C_{\gamma\delta}(\ell)$ assuming the Born-approximation (thick solid line) and corrections with a perturbed geodesic (thin solid lines) for $\gamma,\delta\in\left\{0,2,3,5\right\}$. The background galaxies are all placed at a redshift of $z_s = 0.9$, and the cosmic variance error $\Delta C_{\gamma\delta}(\ell)$ is plotted for comparison (thick dashed line).}
\label{fig_spectra_nonzero}
\end{figure}

\begin{figure}
\begin{center}
\resizebox{0.5\hsize}{!}{\includegraphics{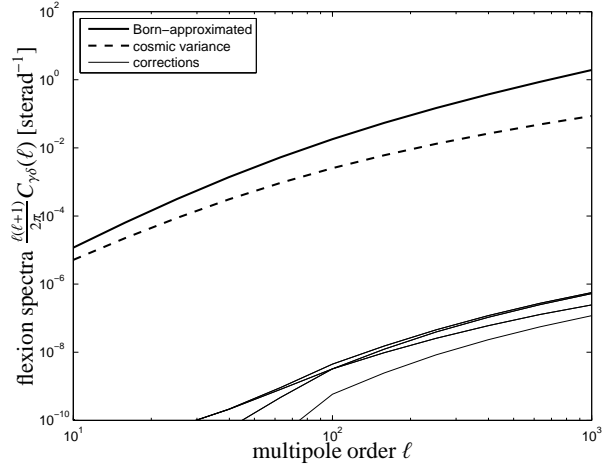}}
\end{center}
\caption{Angular flexion power spectra $C_{\gamma\delta}(\ell)$ involving $\gamma, \delta\in\left\{1,4\right\}$  (thin solid lines), which vanish if the Born-appoximation is applied, but are non-zero from our calculation using a perturbed geodesic. We plot for comparison the flexion spectrum $C_{00}(\ell)$ (thick solid line) with its cosmic variance error $\Delta C_{00}(\ell)$ (thick dashed line) assuming the Born-approximation (compare Fig.~\ref{fig_spectra_nonzero}). The background galaxies are assumed to reside at a redshift of $z_s = 0.9$.}
\label{fig_spectra_zero}
\end{figure}

The spectra of the flexion coefficients are shown in Figs.~\ref{fig_spectra_nonzero} and~\ref{fig_spectra_zero}, where the lensed background galaxies are assumed to reside at a redshift of $z_s=0.9$ ($\sim2.2~\mathrm{Gpc}/h$ in $\Lambda$CDM), which corresponds to the median redshift of the EUCLID galaxy sample. The corrections to the spectra comprise contributions due to the dropped Born-approximation, lens-lens and lens-flexion couplings, to linear order in the perturbed geodesic. If all approximations are in place, the spectra involving decompositions with $\Delta_1$ and $\Delta_4$ are equal to zero, and all other spectra are identical. Corrections arising if the weak flexion signal is evaluated along the true geodesic instead of the Born-approximated geodesic amount to $10^{-4}$ on the smallest angular scales considered, with the difference increasing towards larger angular scales. Nonlinear corrections to the CDM spectrum $P(k)$ on small spatial scales would increase the amplitude of the power by approximately 1.5 dex \citep{2003MNRAS.341.1311S}, and increase the lensing correction amplitude by 3 dex at most (since the corrections are proportional to the square of the power). Hence the relative correction is increased from $10^{-4}$ (in the linear regime) to $3\times 10^{-3}$ in the very nonlinear regime; thus we conclude that for cosmic flexions, the Born-approximation is fulfilled to a very high degree of accuracy. This is despite the fact that there are more contributing terms in the flexion correction in comparison to shear correction; i.e. we find that flexion correction terms are proportionately somewhat smaller than the shear corrections terms already calculated \citep{2006JCAP...03..007S,2002ApJ...574...19C}. 

Note that on very small scales one starts to probe the internal structure of haloes instead of the smooth Gaussian fluctuations of the cosmic density field. In this case, it would be more appropriate to replace the CDM spectrum $P(k)$ with a description following the halo-model \citep[for a review on the halo model including its application to weak lensing, see][]{2002PhR...372....1C}. The more accessible and possibly more relevant quantity would be the 1-halo term, for which one would need to compute the above discussed correction in the case of lensing on a single, spherically symmetric matter distribution. The lens-lens couplings, however, would be difficult to incorporate into this model due to the correlatedness of the individual deflecting haloes along the line of sight.

Comparing the magnitude of the corrections to the uncertainty due to cosmic variance,
\begin{equation}
\Delta C_{\gamma\delta}(\ell) = \sqrt{\frac{2}{2\ell+1}\frac{1}{f_\mathrm{sky}}}C_{\gamma\delta}(\ell),
\end{equation} 
one sees that, for a Euclid-like survey with $f_\mathrm{sky}=1/2$, the corrections are much smaller than cosmic variance for the range of multipoles considered here. Note that the newly excited twist degrees of freedom associated with the decomposition with $\Delta_1$ and $\Delta_4$, reach amplitudes in their power spectra which are slightly smaller than the corrections to flexion power spectra. Hence twist from a cosmic origin will be difficult to detect; if it is present in surveys, it is likely that an origin from systematic effects will dominate.

\section{Conclusion}\label{sect_summary}
This paper treats corrections to the angular spectra of weak cosmic flexions due to nonlocal interactions of the light ray with the cosmic large scale structure, which arise if the Born-approximation is dropped and if lens-lens and lens-flexion couplings are active. We have derived corrections to the spectra by perturbing the geodesic to first order. All integrals needed for computing the geodesic corrections can be represented in a graphical way, which we have summarised in Figure~\ref{fig_comic}. For illustration, we have carried out the calculations for a $\Lambda$CDM cosmology for the planned EUCLID weak lensing survey, which reaches a median redshift of 0.9.

The corrections violate the symmetry of the Jacobian $\mathcal{A}$ and the derivative $\mathcal{B}$ due to their nonlocal nature. Concerning the flexion signal, these fall into three categories: the perturbed geodesic makes the light ray experience gradients in the gravitational potentials at a different position relative to a fiducial straight ray, and solving the implicit propagation equations for $\mathcal{A}$ and $\mathcal{B}$ with a perturbation series give rise to lens-lens couplings and lens-flexion couplings, where the integrated lensing signal is coupled to a derivative of the gravitational potential.

We have explained how the perturbative expansion of the line-of-sight integrals is a perturbation series in the gravitational potential. Powers in the gravitational potential give rise to mode-couplings in Fourier-space when computing angular spectra. Naturally, the mode coupling is strongest between adjacent Fourier-modes and drops rapidly with increasing distance in Fourier space.

We have decomposed the spectra in terms of Pauli- and Dirac-matrices, respectively, which allows us to separate the newly arising distortion modes, and to identify the vectorial flexion. We have derived the perturbative corrections to all spectra of the flexion components, and have shown that they are smaller than the Born-approximated spectrum by approximately four orders of magnitude, well below the cosmic variance limit. The spectra of the newly arising twist components are of similar magnitude compared to the corrections, albeit slightly weaker. All corrections are of similar relative size compared to analogous corrections to weak cosmic shear spectra.

For the case of weak cosmic flexions, we conclude that the Born-approximation remains an excellent approximation. We expect that geodesic corrections are larger in the case of cluster lensing, as the second derivatives of a cluster lensing potential are stronger than those arising in the large-scale structure. It remains to quantify the importance of the flexion corrections in that context, together with the amplitude of the newly excited twist-degrees of freedom associated with $\Delta_1$ and $\Delta_4$.

\section*{Acknowledgements}
BMS's work is supported by the German Research Foundation (DFG) within the framework of the excellence initiative through the Heidelberg Graduate School of Fundamental Physics. LH receives funding from the Swiss science foundation. AFK is funded by GSFP/Heidelberg. \david{DJB is supported by STFC grants ST/H002774/1, ST/F002335/1 and an RCUK Academic Fellowship.}  We would like to thank Matthias Bartelmann, Rob Crittenden, Peter Melchior, Philipp Merkel, Julian Merten and Charles Shapiro for their suggestions.

\bibliography{bibtex/aamnem,bibtex/references}
\bibliographystyle{mn2e}

\bsp

\label{lastpage}

\end{document}